\documentstyle[preprint,aps]{revtex}
\begin{document}
\title{\hfill {\normalsize hepth/9510227}\\[-14pt]
\hfill {\normalsize UCSBTH-95-30}\\[-16pt]
\hfill {\normalsize NSF-ITP-95-136}\\
\smallskip
New Vacua for Type II String Theory}
\author{Joseph Polchinski }
\vskip .13 in
\address{Institute for Theoretical Physics, University of California,\\
Santa Barbara, CA 93106-4030}
\author{and}
\author{Andrew Strominger}
\vskip .13 in
\address{Department of Physics, University of California,\\
Santa Barbara, CA 93106-9530}


\maketitle

\begin{abstract}
\baselineskip=18pt
Lorentz-invariant expectation values for antisymmetric tensor field strengths
in Calabi-Yau compactification of IIA string theory are considered.
These are found to impart magnetic and/or electric charges to the
dilaton hypermultiplet. This results in a potential which can have
supersymmetric minima at zero coupling or at conifold points in the
moduli space. The latter occurs whenever the dilaton charge is aligned
with that of the light black hole at the conifold. It is shown that
there is a flat direction extending from the conifold along which there
is a black hole condensate whose strength is of order the string
coupling $g_s$. It is speculated that these new vacua correspond to
string compactification on generalized Calabi-Yau spaces which have
$c_1=0$ but are not Kahler.

\end{abstract}

\setlength{\baselineskip}{.3in}

\narrowtext

\section{IIA Compactification with Ramond-Ramond Backgrounds}

Consider IIA string theory compactified to four dimensions on a
Calabi-Yau space $X$. The low-energy effective field theory contains
$h_{11} (X)$ vector multiplets with moduli space $M_V$ (which includes
the radial mode) and $h_{21}(X) +1$ neutral hypermultiplets (which
includes the dilaton).

In ten dimensions the IIA theory contains 2-form $(G)$, 4-form $(F)$, and
10-form $(E)$ (as recently discovered in \cite{jppd}) field strengths, as
well as their 8-form, 6-form, and 0-form duals. In this paper we
consider the effects of expectation values for these fields. An
expectation value for the 10-form $E$
\begin{equation}
\langle E\rangle = \nu_0 {\epsilon}^{(10)}
\label{expe}
\end{equation}
where $\nu_0$ is a  constant and ${\epsilon}^{(10)}$ is the
10-dimensional volume element, leads \cite{jppd} to the massive
10-dimensional IIA theory with a cosmological constant equal to
$\frac{1}{2} \nu_0^2$ discovered by Romans \cite{romans}.
In the context of Calabi-Yau compactification, there are $2h_{11}+1$
additional possibilities consistent with four-dimensional Lorentz
invariance. The field strength $G$ may acquire an expectation value
proportional to a harmonic form
$\omega^{(2)}_I$ on $X$
\begin{equation}
\langle G\rangle = \nu^i_2\ \omega^{(2)}_i \quad i=1, \cdots, h_{11}
\ . \label{expg}
\end{equation}
Similarly $F$ may acquire an expectation value proportional to a harmonic
4-form on $X$ or to the 4-dimensional spacetime volume element ${\cal
E}^{(4)}$
\begin{equation}
\langle F\rangle = \nu^i_4\ \omega^{(4)}_i + \nu_6\ {\epsilon}^{(4)}
\quad i=1,\cdots, h_{11}\ . \label{expf}
\end{equation}

Supersymmetry requires that all the $\nu$'s vanish in a Calabi-Yau
vacuum \cite{cand}. Since at the classical level the $\nu$'s can be
continuously adjusted to zero, their effects can be summarized in a
supersymmetric, 4-dimensional effective action in which the $\nu$'s
appear as coupling constants.  Since non-zero values of $\nu$
generically break supersymmetry,\footnote{Quantization
of the $\nu$'s will be discussed below.}  we expect a
moduli-dependent scalar potential ${\cal V}$.

Let us first consider the case where all the $\nu$'s except $\nu_0$
vanish. The resulting 4-dimensional theory is then obtained by
compactification of the massive IIA theory in 10 dimensions. This latter
theory contains a coupling, in the 10-dimensional Einstein frame,
\begin{equation}
\frac{\nu_0}{2} \int d^{10} x \sqrt{-g}\ e^{3\hat\phi/2} G^{MN} B_{NM}
{}~~~~~M, N =0, 1, \cdots, 9
\label{tdc}
\end{equation}
where $B$ is the NS-NS two form potential and the 10-dimensional
string coupling
is $g_s=e^{\hat\phi}$.  The reduction of this includes,
in the four dimensional Einstein frame,
\begin{equation}
\frac{\nu_0}{2} \int d^4 x\ \sqrt{-g}\ e^{ 3D} G^{\mu\nu} B_{\mu\nu}
{}~~~~~
\ \mu, \nu = 0, 1, \cdots, 3\ ,
\label{fdc}
\end{equation}
where the scalar $D$ is related to the string frame volume $V$ of the
Calabi-Yau by
\begin{equation}
e^{3D} = V\ .
\label{dv}
\end{equation}
The field $D$ is an element of a vector multiplet.

The four-dimensional dilaton $\phi$ is part of a neutral hypermultiplet
and is given by
\begin{equation}
\phi = \hat\phi - {3 D\over 2} .
\label{ppd}
\end{equation}
The other three scalars in the hypermultiplet are the NS-NS axion $a$
defined by
\begin{equation}
da = e^{-4\phi}* dB
\label{eight}
\end{equation}
together with the two R-R axions parameterizing the
expectation values  of the 3-form potentials proportional to holomorphic
and anti-holomorphic 3-forms on $X$.

To express (\ref{fdc}) in a manner compatible with four-dimensional
supersymmetry, we define a magnetic potential for $G$
\begin{equation}
e^{3D}*G= d\tilde A\ .
\label{gdp}
\end{equation}
Integrating by parts, (\ref{fdc}) then becomes
\begin{equation}
\nu_0 \int d^4 x\ \sqrt{-g}\ e^{4\phi}\tilde A_\mu \nabla^\mu a\ .
\label{aac}
\end{equation}
This appears to violate (magnetic) gauge invariance, under which
\begin{equation}
\tilde A_\mu\to \tilde A_\mu + \partial_\mu\ \epsilon\ .
\label{ggt}
\end{equation}
However, further couplings complete (\ref{aac}) to the perfect square
\begin{equation}
\frac{1}{2}\int d^4 x\ \sqrt{-g}\ e^{4\phi}g^{\mu\nu}\left(\nabla_\mu a +
\nu_0
\tilde A_\mu\right)\ \left(\nabla_\nu a + \nu_0 \tilde A_\nu\right)\ .
\label{psg}
\end{equation}
Gauge invariance is then restored by accompanying (\ref{ggt}) with the
shift
\begin{equation}
a\to a- \epsilon \nu_0.
\label{aggt}
\end{equation}

The complex scalar
\begin{equation}
S = e^{-2\phi} +ia
\label{sdef}
\end{equation}
which appears in an $N=1$ superfield can then be used to construct an
operator $e^{\lambda S}$ which transforms as
\begin{equation}
e^{\lambda {S}} \to e^{-i\nu_0\lambda\epsilon} e^{\lambda {S}}\ .
\label{strans}
\end{equation}
Thus this operator carries magnetic charge under the $U(1)$
Ramond-Ramond gauge symmetry of the IIA string.  When $e^{\lambda {S}}$
has a nonzero expectation value, this implies that particles carrying
the corresponding electric charge are confined.  This can also be seen
from the fact that the operator~(\ref{fdc}) gives mass to the gauge
field and to $B_{\mu\nu}$.  Noting that the $B_{\mu\nu}$ kinetic
term is proportional to
$e^{-2\phi}$ and that the gauge field kinetic term is not, this mass
is of order $e^\phi$ times the string scale.

One may also consider the effects of the other types of expectation
values. It is not hard to see that for general values the dilaton
acquires magnetic charges $(\nu_0, \nu^i_2)$ and  electric charges
$(\nu_6, \nu^j_4)$. Thus, by turning on R-R backgrounds arbitrary
electric and magnetic charges may be imparted to the dilaton.
NS-NS backgrounds also lead to charges.

There is also a mirror description of this phenomena in the
context of IIB string
theory. In that case there are two 3-form field strengths
(one NS-NS and one R-R) which can
acquire expectation values proportional to harmonic 3-forms on the
Calabi-Yau.

\section{Quantization of the Field Strength}

The fact that $e^{\lambda {S}}$ gets a magnetic charge proportional to
$\nu_0$ strongly suggests that the latter is quantized.
Ramond-Ramond charge is believed to be quantized in a unit which was
denoted $\mu_0$ in ref.~\cite{jppd}.  Then the corresponding magnetic
charge is quantized in units of $2\pi/\mu_0$, and so $\epsilon \sim
\epsilon + \mu_0$ and the gauge transformation implies that\footnote
{We have normalized the R-R fields to have a canonical kinetic
term ($\alpha_p = 1$ in the notation of ref.~\cite{jppd}), and
the NS-NS field $B_{\mu\nu}$ such that its
coupling to the fundamental string is $(2\pi\alpha')^{-1} \int B$.}
\begin{equation}
a \sim a + \nu_0 \mu_0.
\label{fourteen}
\end{equation}
However, $a$ was already periodically identified because of the shift
encountered in encircling a fundamental string
\begin{equation}
a \sim a + \frac{1}{2\pi\alpha'}
\label{astring}
\end{equation}
These identifications must be commensurate, so $\nu_0$ is
quantized.  If, as seems likely, the identifications are the same, then
\begin{equation}
\nu_0\mu_0 = \frac{1}{2\pi\alpha'}.
\label{nuquant}
\end{equation}
Using the calculation of the charge quanta from ref.~\cite{jppd}, this
implies that $\nu_0$ is exactly $\mu_8$, the charge carried by the
8-brane. In other words, all values of $\nu_0$ that are allowed by the
quantization are dynamically accessible by nucleation of 8-branes.

It is interesting to derive the same result directly in ten dimensions.
Consider an 8-sphere surrounding a 0-brane which carries Ramond-Ramond
charge.  Integrating the field equation
\begin{equation}
d*(e^{-2\phi} H / 2) =\nu_0 *(G+\nu_0 B)
\end{equation}
over the 8-sphere seems to imply that $0 = \nu_0 \int *(G+\nu_0 B) $,
so the total
flux is zero if $\nu_0 \neq 0$.  This would say that such charge is not
observable at any scale, a result which is surprising because we have
seen in the preceding section that the confinement scale is somewhat
below the string scale.  Suppose, however, that a fundamental string
ends on the 0-brane, a possibility suggested by the identification of
the latter as a D-brane.  This adds a source
$(2\pi\alpha')^{-1}$ times a delta-function to the field equation, and
since the string passes through the 8-sphere once we now have
$0 = \nu_0 \int *(G+\nu_0 B)  + (2\pi\alpha')^{-1}$, giving the
same quantization
found above.

Thus, we have an interesting physical picture.  Suppose that $\nu_0 = n
\mu_8$, and that there is a 0-brane/anti-0-brane pair each with a
single unit of charge.  Then the $G$-flux runs between them in a tube of
width $g_s^{-1}$ times the string scale, and in addition they are
connected by $n$ fundamental strings.

It is possible that this confinement will play an interesting role in
string phenomenology, for example by removing states from the spectrum.

\section{The Scalar Potential at Large Radius}

The scalar potential ${\cal V}$ in four dimensions is determined by the
charges of
the hypermultiplets and by $N=2$ supersymmetry. The general form of the
potential is given for the case of electric charges in
refs.~\cite{ferrara,dlwp}.  This can be applied to the present
(magnetically charged) case by simply performing a symplectic
transformation which trades $G$ for $*G$ and turns the magnetic charge
into an electric one
\cite{CDFV}. The electric formula is
\begin{equation}
{\cal V} = h_{uv} k^u_I k^v_J \bar X^I X^J e^K +
(U^{IJ} - 3\bar X^I X^J e^K) {\cal P}^i_I {\cal P}^i_J\ .
\label{vdef}
\end{equation}
On the vector multiplet moduli space $M_V$, $X^I$ for $I=0, 1, \cdots,
h_{11}$ are complex projective coordinates, $K$ is the Kahler potential,
which is related to a choice of holomorphic sections $(X^I, F_I)$ by
$K = -\ln i(\bar X^I F_I - X^I \bar F_I)$,
and
\begin{equation}
U^{IJ} =e^K(\partial_a+\partial_aK)X^Ig^{a\bar b}(\partial_{\bar b}+
\partial_{\bar b}K)\bar X^{J}
\label{eighteen}
\end{equation}
with $a,~b$ (non-projective) tangent space indices.  On the
hypermultiplet moduli space $M_H$,
$h_{uv}$ is the metric,
$k^u_I$ is the Killing vector which generates the action of the gauge
transformation for the $I$'th abelian gauge field, and ${\cal P}^i_I$ for
$i=1,2,3$ are an $SU(2)$ triplet of Killing potentials.

We will be concerned with the case that only one charge, which shall be
labeled ``0,'' is non-vanishing.  The non-zero components of $k$ follow
simply from (\ref{aggt}) as
\begin{equation}
k^u_0 \frac{\partial}{\partial u} = \nu_0 \frac{\partial}{\partial a} \ .
\label{kva}
\end{equation}
In this section the only relevant hypermultiplet is that containing
the dilaton.  The corresponding quaternionic geometry is
obtained from ref.~\cite{fersab} by taking the special case $n = 0$.  In
terms of the $N=1$ dilaton superfield $S$ and a superfield $C$ for the
R-R sector components, the moduli space is
Kahler with $K_H = -\ln 2(S + \bar S - [C+\bar C]^2)$.
The norm of $k$ is then
\begin{equation}
k^u_0 k^v_0 h_{uv} = \frac{\nu_0^2}{4}e^{4\phi}\
\label{knr}
\end{equation}
where, for nonzero $C$, we define the dilaton by $2 e^{-2\phi} = S + \bar S
- [C+\bar C]^2$.
Define
\begin{equation}
u = e^\phi dC, \qquad v = e^{2\phi} [dS/2 - (C+ \bar C) dC],
\label{uv}
\end{equation}
Let $e = (u,v)$.  Then the form $\Omega^i$~\cite{ferrara} constructed
from the triplet of complex structures, and the corresponding potential
$\omega^i$, are found to be
\begin{eqnarray}
&&\Omega^i = i e^\dagger \sigma^i e \nonumber\\
&&\omega^1 = i(\bar u - u), \qquad \omega^2 = (\bar u + u), \qquad
\omega^3 =i (v-\bar v)/2
\end{eqnarray}
These are related by
\begin{equation}
d \Omega^i + \epsilon^{ijk} \omega^j\wedge \Omega^k=0.
\label{wwterms}
\end{equation}
The Killing potentials ($D$-terms) are derived from the relation
\begin{equation}
{\bf i}_{k_I} \Omega^i = - d {\cal P}_I^i - \epsilon^{ijk} \omega^j {\cal
P}_I^k
\label{dterms}
\end{equation}
(${\bf i}_{k}$ denoting contraction of a form with a vector).
The result is that the only nonvanishing component is
\begin{equation}
{\cal P}_0^3 =-
\frac{1}{2}\nu_0 e^{2\phi}.
\label{dtms}
\end{equation}

At large volume the only relevant vector multiplet is that containing
the radial mode and corresponding axion, $t = b + ie^D$.  The usual
choice
of sections is
\begin{equation}
\widetilde\Pi = \pmatrix{\widetilde F_0\cr \widetilde F_1\cr
\widetilde X^0\cr \widetilde X^1\cr}
\buildrel{D\to\infty}\over= \pmatrix{5t^3/6 \cr -5t^2/2 \cr 1 \cr t\cr},
\label{pv1}
\end{equation}
but because the 0 charge is magnetic we must make a symplectic
transformation to
\begin{equation}
\widetilde\Pi' = \pmatrix{\widetilde F_0'\cr \widetilde F_1'\cr
\widetilde X'^0\cr \widetilde X'^1\cr}
 = \pmatrix{\widetilde X_0\cr \widetilde F_1\cr -\widetilde F_0 \cr
\widetilde X^1\cr}\ .
\end{equation}
in order to apply the formulae of \cite{CDFV}.
One then finds that all terms in the potential~(\ref{vdef}) scale as
\begin{equation}
{\cal V}\ \buildrel{D\to\infty}\over\sim\ \nu^2_0 e^{4\phi + 3D}
\label{rgv}
\end{equation}
in agreement with the potential derived by direct reduction of the
10-dimensional cosmological constant in the massive IIA theory.
Similarly, an electric charge gives a potential $\sim \nu^2_6 e^{4\phi -
3D}$, again in agreement with that found directly from the action for
$F$.

\section{$N=1$ Supersymmetry}

Although we are considering here a compactification with $N=2$
spacetime supersymmetry, it is worth noting that the same phenomena
will arise in $N=1$ compactifications, and the potential is easily
derived using $N=1$ superfields.  The difference from the familiar
heterotic string compactifications is that the function $f$ multiplying
the gauge field kinetic terms is now independent of the dilaton, since
the gauge fields are in the R-R sector.  For the gauge field
which descends from the 10-dimensional gauge field, the kinetic term
is proportional to the volume of the compactified space, so $f \sim
t^3$.  The Kahler potential for the dilaton is the usual $-\ln(S + \bar
S)$.  When the dilaton is charged the $D$-terms give a potential
\begin{equation}
{\cal V} = \frac{1}{{\rm Im}(f)} \delta K \bar \delta K,
\end{equation}
where $\delta$ and $\bar \delta$ refer to taking gauge variations of
the chiral superfields and their conjugates respectively.
An electric charge on the dilaton gives rise to a
potential $S^{-2} t^{-3}$.  For a magnetic charge we must first make a
symplectic transformation to $f \ \propto\ t^{-3}$, and so the
potential is $S^{-2} t^3$.  These are the same as found above.

Related mechanisms were discussed in~\cite{dsw}.  That paper
considered the heterotic string, which has the important difference
that a charge on the dilaton leads to an anomalous variation of the
action, from the dependence of $f$ on $S$.  Thus the charges carried by
the dilaton are fixed so as to cancel the $U(1)$ anomalies
from light fields.  In the type~II string the charge on the dilaton is
not constrained by the anomaly.  Also considered in
refs.~\cite{dsww,dsw} were similar charges on moduli fields.  These are
not constrained by the anomaly, but they cannot appear in the low energy
theory of the heterotic string because they lead to a potential for the
moduli which is of order the string scale~\cite{dsww,dsw}.  By
contrast, the effect of the R-R field strengths can be
regarded as a perturbation.  This can be seen from the fact
that the cosmological constant
vanishes as the coupling is taken to
zero in either the string or Einstein frame.

It will be interesting to find the heterotic string duals of our type II
backgrounds.  We have not yet explored this issue, except to note
that the background $\nu_6$ becomes under six-dimensional string-string
duality a magnetic field in the 4-5 direction.  This heterotic
background has been considered in the interesting
recent work~\cite{bachas} and was also one
motivation for ref.~\cite{GPS}.  It corresponds to giving a charge to a
modulus~\cite{dsww,dsw}.

\section{The Scalar Potential Near a Conifold}

Typically, as in (\ref{rgv}), the potential will drive the theory to
zero coupling which is uninteresting, or towards strong coupling, where
non-perturbative effects must be considered.\footnote
{However, we should note the recent result~\cite{FGP} that electric and
magnetic field strengths in different $U(1)$'s can break $N=2$
supersymmetry to $N=1$.}
Life becomes much more interesting when one includes the effects of the
charged BPS black holes (which can also be represented as
D-branes~\cite{jppd}). In the IIA theory these arise from 0-, 2-, 4-, and
6-branes which wrap minimal, supersymmetric cycles in
$X$. These can become massless at conifold points in the moduli space
where the (world-sheet-instanton-corrected) periods degenerate.  One
then finds that the potential has flat directions corresponding to new
supersymmetric string vacua.

To understand this in detail, let us consider the specific example of
the IIA theory on the quintic $P_4(5)$ with $\nu_0\not= 0$ so that
the dilaton carries magnetic charge. This theory has a single vector
multiplet $(h_{11} =1)$  which includes the radial mode $D$. The moduli
space $M_V$ has a conifold singularity \cite{cgpo} at which a single
BPS black hole degenerates to zero mass \cite{mbh}. It turns out that
this black hole corresponds to a 6-brane wrapping the entire Calabi-Yau
and carries the same magnetic charge as the dilaton. To see this we note
that the geometry of $M_V$ is usually described by a period vector
\begin{equation}
\Pi = \pmatrix{F_0\cr F_1\cr X^0\cr X^1\cr}
\label{pv}
\end{equation}
adapted to the mirror of $P_4(5)$ in which the entries are 3-cycle
periods of the holomorphic 3-form $\Omega$. A conifold singularity
occurs at a point when $X^1=0$ and $F_1(X^1) \sim
\frac{1}{2\pi i} X^1 \ell n X^1$. This is related to the earlier
basis~(\ref{pv1}) by a symplectic transformation (for
some choice of Kahler gauge for $\Pi$) which was determined in
\cite{cgpo} as
\begin{equation}
\widetilde\Pi = N\, \Pi
\label{ppdr}
\end{equation}
where
\begin{equation}
N=\pmatrix{0&0&0&1\cr
           -1&0&0&0\cr
           0&-1&0&0\cr
           -2&0&-1&0\cr}\ .
\label{nmat}
\end{equation}
\vskip .13 in
\noindent
Hence if the period $X^1$ of $\Omega$ on the mirror of $P_4(5)$ is
degenerating, the period $\widetilde F_0$, which approaches the volume at
large radius, is also degenerating. It may sound rather strange that the
entire 6-volume of $P_4(5)$ degenerates (while other cycles remain
finite), but this is in a region where worldsheet instanton corrections
are large and the distinction between 0-, 2-,4-, and 6-cycles is lost.

In ten dimensions, the 6-brane acts as a source for the 8-form $*G$.
Hence a 6-brane which wraps the Calabi-Yau carries the magnetic charge
associated to $G$, just like a dilaton which acquires charge due to
nonzero $\nu_0$. In four dimensions this leads to a charged
hypermultiplet which we will represent by an $SU(2)$ doublet $B$.

Next, let us consider the dilaton potential on the moduli space $M_V
\otimes M_H$, ignoring for a moment the field $B$. It is convenient to
work in the $\Pi$ basis (\ref{pv}), in which the dilaton carries
electric charge in the ``1'' direction. The nonzero component of the
Killing vector is
then
\begin{equation}
k_1^u \frac{\partial}{\partial u} = \nu_0 \frac{\partial}{\partial a}\ .
\label{kfrm}
\end{equation}
Near the conifold at $X^1=0$, $g_{1\bar 1}$ diverges as
\begin{equation}
g_{1\bar 1} \sim -\ln( \bar X^1 X^1)
\label{ncnf}
\end{equation}
It follows that near the conifold, ${\cal V}$ has a local minimum
\begin{equation}
{\cal V}\sim -{\nu_0^2 \over \ln( \bar X^1 X^1)}\ .
\label{thirtyone}
\end{equation}
Hence, as pointed out in \cite{mbh}, the effective field theory without the
black
hole field $B$ can be used to show that the moduli are attracted to the
conifold.

Near the conifold, the $B$ field gets light and must be
included in the effective field theory. To leading order in $B$ and
$e^\phi$,  (\ref{kfrm}) becomes
\begin{equation}
k^u_1\frac{\partial}{\partial u} =  \nu_0 \frac{\partial}{\partial a}
+ B\ \frac{\partial}{\partial B} -  B^\dagger
\ \frac{\partial}{\partial  B^\dagger}
\label{kpb}
\end{equation}
where $B$ is a complex doublet. The Killing potential is
\begin{equation}
{\cal P}^i_1 = -\frac{\nu_0}{2} e^{2\phi} \delta^{i3}  + B^\dagger
\sigma^i B\ .
\label{dpb}
\end{equation}
The metric $g$ is now non-singular, the singularity in
(\ref{ncnf}) having arisen from integrating out $B$ \cite{mbh}.  The
first term in the potential~(\ref{vdef}) vanishes at the conifold
point~$X^1 = 0$, and the potential has a flat direction with
unbroken $N=2$ supersymmetry, characterized by
\begin{equation}
{\cal P}^i_1 = 0\ .
\label{}
\end{equation}
The vacua are parameterized by a
single hypermultiplet which is a linear combination of $\Phi$ and $B$.
At generic points the string coupling is non-zero and there is a black
hole condensate. The vector modulus acquires a mass and is frozen at the
conifold.

An important feature of these new vacua is that the expectation value of
the black hole field is of order $g_s = e^\phi$. By taking $g_s$ to be
small we can stay within the validity of both string perturbation theory
and the expansion of the black hole hypermultiplet geometry (known only to
leading order in $B$) about the conifold. The origin of this happy circumstance
is ultimately the fact that the R-R fields appear with an extra factor
of $g_s$ in the supersymmetry transformation laws.

In \cite{gms} it was argued that in some cases branches of the vacuum
moduli space with black hole condensates were a dual description of
known Calabi-Yau compactifications. This is unlikely to be the case here
since this compactification has no massless vector multiplets and so
cannot be a IIA Calabi-Yau compactification. A more likely possibility
is that it corresponds to string compactification on one of the
generalized (non-Kahler) Calabi-Yau spaces obtained by resolutions of
small 3-cycles as 2-cycles. Such spaces are discussed for example in
\cite{reid,cgpo}.  They are complex manifolds with $c_1=0$ but are in
general not Kahler.

There are a number of extensions of our observations. Of
particular interest is the possibility of breaking $N=2$
to $N=1$ along the lines recently described in \cite{FGP}.
This appears to require simultaneous expectation values for both
R-R and NS-NS field strengths. Dilaton charges may
also lead to new $N=4$ compactifications.

\acknowledgments

We are grateful to K.~Becker, M.~Becker, S.~Ferrara, J.~Harvey, D.~Lowe,
S.T.~Yau, and especially G.~Moore for useful conversations. A recent
attempt to generate a $D$ term in $N=2$ theories has been made by
J.~Harvey and G.~Moore (unpublished). This work was supported in part by
the Department of Energy, grant \#91ER40618 and NSF grants
PHY91-16964 and PHY94-07194.


\begin{references}

\bibitem{jppd} J.~Polchinski, {\it Dirichlet Branes and Ramond-Ramond
Charges}. preprint NSF-ITP-95-122, hepth/9510017 (1995).

\bibitem{romans} L.~Romans, {\sl Phys.~Lett.}, {\bf B169}, 374, (1986).

\bibitem{cand} P.~Candelas and D. Raine, {\sl Nucl.~Phys.}, {\bf B248}, 415,
(1984).

\bibitem{ferrara} R.~d'Auria, S.~Ferrara, and P.~Fre, {\sl Nucl.~Phys.},
{\bf B359}, 705 (1991).

\bibitem{dlwp} B.~deWit, P.~Lauwers, and A.~van Proeyen, {\sl
Nucl.~Phys.}, {\bf B255}, 569 (1985).

\bibitem{CDFV} A. Ceresole, R. D'Auria, S. Ferrara, and A. Van
Proeyen, {\sl Nucl. Phys.} {\bf B444}, 92 (1995).

\bibitem{fersab} S. Ferrara and S. Sabharwal, {\sl Nucl. Phys.}
{\bf B332}, 317 (1990).

\bibitem{dsw} M.~Dine, N.~Seiberg, and E.~Witten,
{\sl Nucl. Phys.} {\bf B289}, 589 (1987).

%

\bibitem{dsww} M. Dine, N. Seiberg, X.G. Wen and E. Witten,
{\sl Nucl. Phys.} {\bf B289}, 319 (1987).

\bibitem{bachas} C. Bachas, {\it A Way to Break Supersymmetry,} preprint
CPTH-R349-0395, hep-th/9503030 (1995).

\bibitem{GPS}  S. B. Giddings, J. Polchinski, and A. Strominger,
{\sl Phys. Rev.} {\bf D48}, 5784 (1993).

\bibitem{FGP}
S. Ferrara, L. Girardello, M. Porrati,
preprint NYU-TH-95-10-02, hep-th/9510074 (1995).

\bibitem{cgpo} P.~Candelas, L.~Parkes, P.~Green, and X.~de la Ossa,
{\sl Nucl. Phys.} {\bf B359}, 21 (1991).

\bibitem{mbh} A.~Strominger,
{\sl Nucl.~Phys.},  {\bf B451}, 96 (1995);  hep-th/9504090.

\bibitem{gms} B.~Greene, D.~Morrison, and A.~Strominger, {\sl
Nucl.~Phys.}, {\bf B451}, 109 (1995).

\bibitem{reid} M.~Reid, {\sl Math.~Ann.}, {\bf 274}, 329 (1987).

\end{references}
\end{document}